\documentclass[
    ,final            
  ]
  {aipproc}

\layoutstyle{6x9}

\def\lp {\left( }
\def\rp {\right) }
\def\lb {\left[ }
\def\rb {\right] }
\def\lc {\left\{ }
\def\rc {\right\} }

\def\nn {\nonumber}

\def\beq{\begin{equation}}
\def\eeq{\end{equation}}
\def\bea{\begin{eqnarray}}
\def\eea{\end{eqnarray}}
\def\ni{\noindent}

\def\di {\partial_\mu }
\def\ds {\partial^\mu }

\def\cd {\!\cdot\!}

\def\rar {\rightarrow}

\def\sp {\!+\!}
\def\sm {\!-\!}

\def\a{\alpha}
\def\b{\beta}

\def\g{\gamma}

\def\G {\Gamma}

\def\m{\mu}

\def\p{\pi}

\def\s{\sigma}

\def\cL {{\cal L}}
\def\cM {{\cal M}}

\def\bpi {\mbox{\boldmath $\pi$}}
\def\bphi {\mbox{\boldmath $\phi$}}

\begin{document}

\title{Chiral Breit-Wigner}



\author{L. O. Arantes and \underline{M. R. Robilotta}}{
  address={Instituto de F\'{\i}sica, Universidade de S\~{a}o Paulo,\\
C.P. 66318, 05315-970, S\~{a}o Paulo, SP, Brazil.}}

\begin{abstract}
Chiral symmetry and unitarization are combined into generalized Breit-Wigner expressions describing 
scalar resonances, which contain free parameters and allow flexible descriptions of masses, widths 
and pole positions.
This theoretical tool is especially designed to be used in analyses of low-energy data.
\end{abstract}

\maketitle



The vacuum of quantum chromodynamics (QCD) is non-trivial and, at low-energies, it can be excited
in the form of pions, kaons or etas.
These particles can be clearly perceived by experiment because their quantum numbers 
allow a neat contrast with the vacuum.
The same kind of contrast is not possible in the case of scalar mesons and their identification
is much more difficult.

Evidences for resonances with low masses and large widths produced by the E791 Fermilab 
experiment in $D$ decays\cite{cbpf} and confirmed in a number of other reactions\cite{OR}
have motivated a renewal in the interest on the scalar sector.
The information extracted from these experiments relies on theoretical expressions 
for resonance widths and amplitudes, which eventually do play major roles in final results.
As a consequence, "empirical" values for masses and coupling constants depend heavily on the 
theoretical input used in data analyses.

\vspace{2mm}
\ni
{\bf CHIRAL SYMMETRY - }
The best theoretical framework available for describing hadronic decays of heavy mesons is 
low-energy QCD, which can be reliably represented by means of effective theories 
with approximate chiral symmetry.
This symmetry is especially suited for describing processes involving pions and has been 
successfully applied to a rather large sample of physical problems.

About ten years ago, T\"ornqvist\cite{Tor} employed chiral symmetry in the description of 
scalar resonances and obtained a width for the decay into $S$-wave pions of the form  

\beq
\G_T(s) = \a \;  (2\, s\sm \m^2) \; \frac{\sqrt{s-4 \m^2 }}{\sqrt{s}}\;e^{-(s-M^2)/4\b^2} \;,
\label{1}
\eeq

\ni
where $\a$ and $\b$ are free parameters and $\m$ and $M$ are the pion and resonance masses.
This result was derived from a quark model and has the merit of including the chiral factor 
$(2\,s \sm \m^2)$, which makes the width small at low-energies.
As we discuss in the sequence, effective lagrangians incorporating chiral symmetry give rise 
to more general expressions, which include other forms of $s$-dependence. 
Another advantage of the lagrangian framework is that it allows the unambiguous separation 
between resonance and background contributions and avoids the risk of double counting.


Modern chiral perturbation theory employs non-linear realizations of the symmetry and
gives rise to results which are very general and incorporate all the possible freedom 
compatible with the symmetry. 
The old and well known linear $\s-$model, which has been much used in the study of 
scalar resonances, can be recovered as a special case. 

\vspace{2mm}
\ni
{\bf RESONANCES AND {$\bpi\bpi$} AMPLITUDE - }
The scalar resonance propagator at low energies is closely related with the elastic
$\p\p$ scattering amplitude, which is the main locus of chiral symmetry.
Quite generally, the amplitude $T_{\p\p}$ for the process 
$ \p(p)\,\p(q) \rar \p(p')\,\p(q')$ 
can be written as 

\vspace{-2mm}

\beq
T_{\p\p} = T_0(s,t,u)\;P_0 +  T_1(s,t,u)\;P_1 +  T_2(s,t,u)\;P_2 \;,  
\label{2}
\eeq

\ni
where $s=(p\!+\! q)^2$, $t=(p\!-\!p')^2$, $u=(p\!-\!q')^2$ and the operator $P_I$ is the 
projector into channel with isospin $I$.

A chiral theorem\cite{Wei66} ensures that the leading order amplitude $T_0$ at low energies
is given by 

\vspace{-3mm}

\beq
T_0 = (2\,s \sm \m^2)/f_\p^2 + \cdots  \;,
\label{3}
\eeq 

\ni
where $\m$ and $f_\p$ are the pion mass and decay constant, whereas the ellipsis indicates 
higher order contributions.
When resonances are included, it is very important to ensure consistency with this theorem.
In the case of a scalar-isoscalar resonance, the tree level amplitude for $\p\p$ scattering 
is given by the four diagrams

\vspace{-79mm}

\begin{figure}[h]
\includegraphics[width=0.78\columnwidth,angle=0]{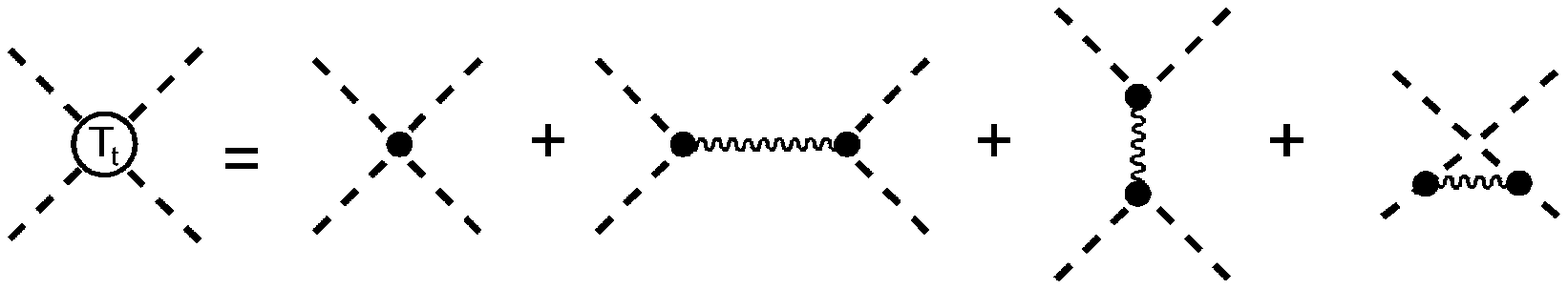}
\end{figure}

\vspace{-70mm}

In the non-linear approach, the resonance field $f$ is assumed to be a chiral scalar 
and couples to pion fields $\bphi$, which behave non-linearly under chiral 
transformations\cite{Wei68}.
The effective lagrangian for this system is written as

\vspace{-1mm}

\bea
\cL &=& \frac{1}{2}\lp \di f  \; \ds f  - M^2 f^2 \rp
+ \frac{1}{2} \lp 1 \sp c_s \,f /f_\p  \rp \lp \di \bphi \cd \ds \bphi 
+\di \sqrt{f_\p^2 \sm \bphi^2}\; \ds \sqrt{f_\p^2 \sm \bphi^2} \rp
\nn\\
&+& \m^2 f_\p \lp 1 \sp c_b \,f / f_\p  \rp \lp \sqrt{f_\p^2 \sm \bphi^2}- f_\p \rp \;,
\label{4}
\eea

\ni
where the dimensionless constants $c_s$ and $c_b$ represent, respectively, the scalar-pion 
couplings that preserve and break chiral symmetry.
The explicit evaluation of the diagrams yields the following scalar-isoscalar tree level  
amplitude

\vspace{-2mm}

\bea
&& T_t = - \; \frac{\g^2}{s-M_\s^2}\;,
\label{5}\\
&& \g^2(s) =  \lc (2\,s \sm \m^2) (M_\s^2 \sm s) + 
3 \lb c_s \, s/2 - (c_s \sm c_b)\, \m^2 \rb^2  + B (M_\s^2 \sm s) \rc  / f_\p^2 \;,
\nonumber
\eea

\ni
where the function $\g(s)$ incorporates chiral dynamics and implements effective 
couplings at the vertices.
Its first term reproduces the low-energy theorem, given by eq.(\ref{3}), 
the second one describes the resonance couplings, and the last function represents a 
non-resonating background, which receives contributions from $t$ and $u$ channels only.
The particular choice $(c_s\!=\!2, c_b\!=\!1)$ corresponds to the linear $\s-$model.


\vspace{2mm}
\ni
{\bf UNITARIZATION - }
The range of validity of the field theoretical $\p\p$ amplitude can be extended 
by means of unitarization.
We consider only iterated contributions from a single loop, without form factors. 
The advantage of this procedure is that there is little model dependence and loop contributions 
are given by a compact analytical expression, which is real below threshold and complex afterwards.
The dressed propagator is determined by the three diagrams shown in figure $(a)$ below,
whereas the full amplitude $T$, described in figure $(b)$, corresponds to a composite Dyson series 
and includes all possible iterations of the $\p\p$ tree amplitude.

\vspace{-75mm}

\begin{figure}[h]
\includegraphics[height=0.80\textheight]{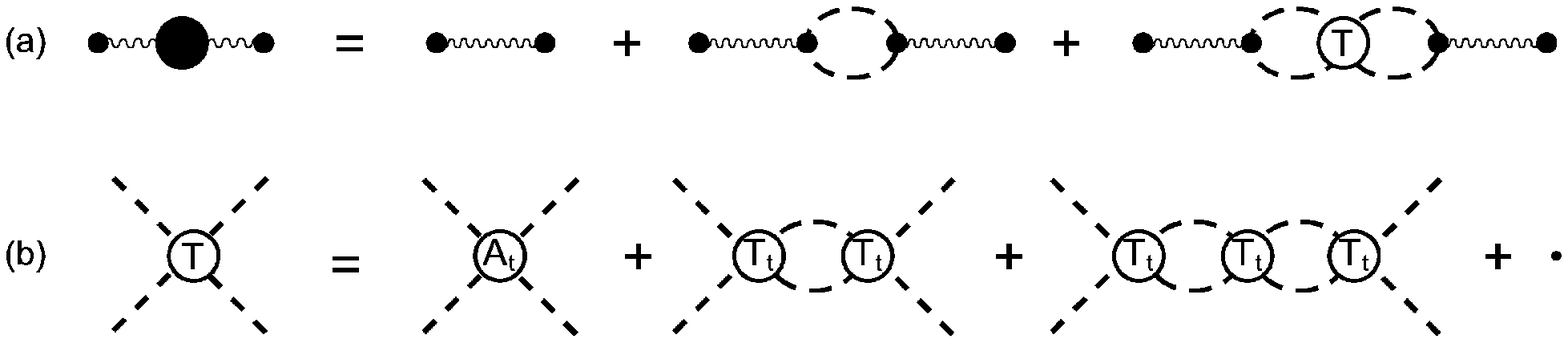}
\end{figure}

\vspace{-67mm}

This geometrical series can be summed and one finds 

\beq 
\bar{T}(s) = -\; \frac{\g^2} {s-\cM^2 + i\;M \;\Gamma_\chi} 
\;\;\;\;\;\;\;
\leftrightarrow
\;\;\;\;\;\;\; 
\G_\chi(s) =  \g^2 \; \frac{\sqrt{s-4\m^2}}{32 \p M \sqrt{s}}\; 
\Theta(s\!-\!4\m^2) \;,
\label{6}
\eeq

\ni
where $\cM^2(s)$ is a running mass such that $\cM^2(M^2)= M^2$.
This result corresponds to the most general expression possible for the width dictated by 
chiral symmetry and should be compared with eq.(\ref{1}).
The signature of the symmetry is the factor $(2 s \sm \m^2)/f_\p^2$ in the the function $\g^2(s)$
[eq.(\ref{5})], which ensures that the low energy theorem for $\p\p$ scattering is satisfied.

\vspace{2mm}
\ni
{\bf RESULTS - }
In the figure below, we explore the interplay between chiral symmetry and resonance in the function 
$|T(s)|^2$, for a scalar mass fixed at $M=\sqrt{12}\, \m \sim$ 485 MeV.
The first term of eq.(\ref{5}) yields the {\em leading order} curve, an unbound parabola 
which blows up at large energies.
The inclusion of the remaining two terms with $(c_s\!=\!2, c_b\!=\!1)$ in that amplitude gives 
rise to the {\em tree} curve.
The {\em unitarized $\s$-model} curve corresponds to eq.(\ref{6}) and is obtained by iterating the 
{\em tree} amplitude by means of two-pion loops.
Finally, the {\em resonance} curve is derived by iterating just the second term in eq.(\ref{5})
with $s=M^2$ and corresponds to the width normally used in data analyses,which is given by 

\beq
\G_R(s) = \frac{3\, (M_\s^2 \sm \m^2)^2 }{f_\p^2}\;
\frac{\sqrt{s-4\m^2}}{32 \p M \sqrt{s}}\;
\Theta(s\!-\!4\m^2) \;.
\label{7}
\eeq
 
Inspecting the figure, one learns that the proper implementation of chiral symmetry forces 
the {\em leading order}, {\em tree} and {\em unitarized $\s$-model} curves to stay very close 
together at low energies.
This kind of behavior also holds for other choices of the parameters 
$(c_s, c_b)$.
In the {\em resonance} curve, on the other hand, the symmetry is badly violated, since it does 
not tend to the {\em leading order} one when $s\rar 0$, as predicted by the low-energy theorems.

\vspace{33mm}

\begin{figure}[h]
\includegraphics[height=0.42\textheight]{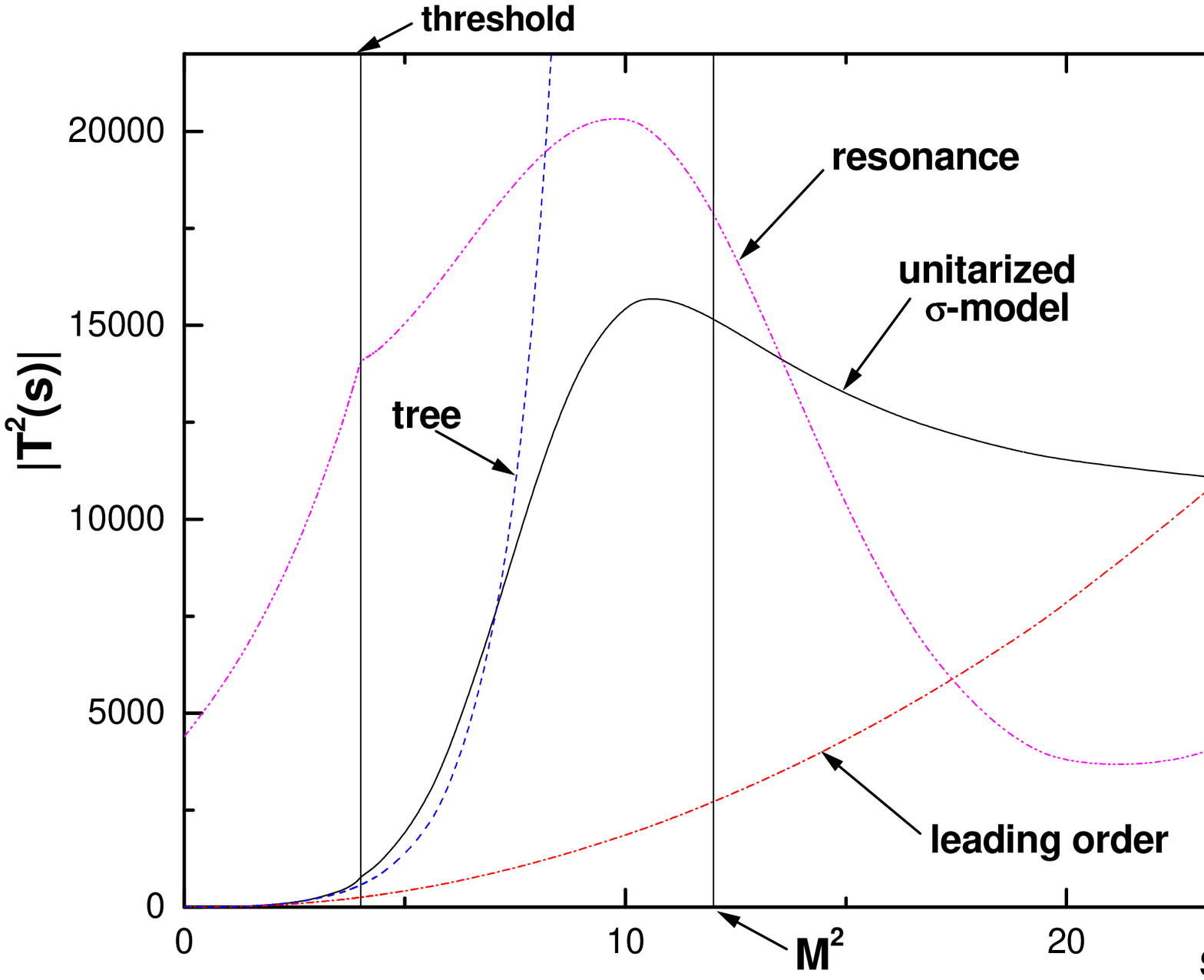}
\end{figure}

\vspace{-47mm}

Resonances manifest themselves as poles in the second Riemann sheet of  scattering amplitudes and
another problem with the use of eq.(\ref{7}) is that it does not allow one to explain how a scalar
mass around $850$ MeV, as observed in phase shifts, can coexist with a pole at much lower energy. 
The strong $s$-dependence of the function $\g$ in the chiral width [eq.(\ref{6})] is essential
to explain\cite{CGL} how scalar parameters $(M_\s, \G_\s/2) \simeq (846, 265) $ MeV
can be compatible with a pole position at $\sqrt{s} \simeq (455 - i\, 290)$ MeV.


\vspace{2mm}
\ni
{\bf CHIRAL BREIT-WIGNER - } Our chiral expression describing a single scalar 
resonance has the following properties:\\[1mm]
$\bullet$ at low energies, it reproduces the theorem derived by Weinberg\cite{Wei66} for $\p\p$ 
scattering;\\[1mm]
$\bullet$ it is compatible with a rather general non-linear realization of chiral symmetry and
involves two free parameters $(c_s, c_b)$, which allow a rather flexible description of masses, 
widths and pole positions as shown in the figure in the next page;\\[1mm]
$\bullet$ results from the usual $K$-matrix unitarization can be recovered by identifying the
running mass in eq.(\ref{6}) with the resonance mass and correspond to the expression 

\vspace{-3mm}

\beq
\bar{T}(s) = -\; \frac{\g^2} {s- M^2 + i\;M \;\Gamma} \;,
\label{8}
\eeq

\ni
with 

\vspace{-8mm}

\bea
&& \G_\chi(s) =  \g^2 \; \frac{\sqrt{s-4\m^2}}{32 \p  M \sqrt{s}}\; \Theta(s\!-\!4\m^2) \;,
\label{9}\\
&& \g(s) =  \lc (2\,s \sm \m^2) (M^2 \sm s) + 3 \lb c_s \, s/2 - (c_s \sm c_b)\, 
\m^2 \rb^2 \rc / f_\p^2 \;.
\label{10}
\eea

\ni
These expressions represent a compromise between simplicity and chiral low-energy dynamics,
which can be used in data analyses by treating the parameters $M$, $c_s$ and $c_b$ 
as being adjustable.

\newpage
.

\vspace{20mm}

\begin{figure}[h]
\includegraphics[height=0.42\textheight]{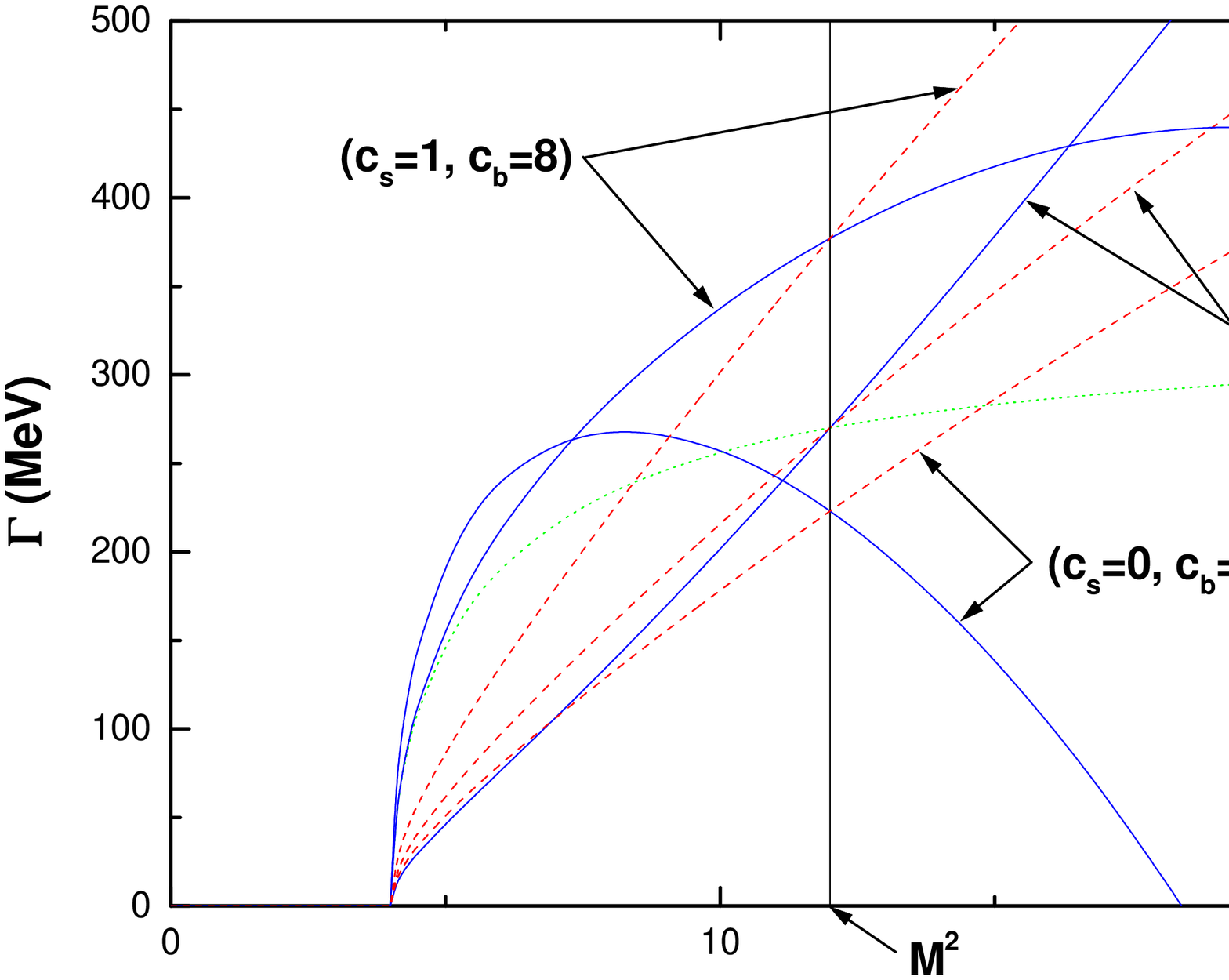}
\end{figure}


\vspace{-46mm}

\ni
$\bullet$ the inclusion of a second resonance with the same quantum numbers is straightforward.
Dynamical unitarization by means of quantum field theory techniques corresponds to 
first combining tree-level amplitudes and subsequently evaluating loop corrections.
This leads to the following trial function:

\beq
\bar{T}(s) = -\, \frac{\g_\a^2\,(s \sm M_\b^2) + \g_\b^2\,(s \sm M_\a^2)}
{(s \sm M_\a^2) \, (s\sm M_\b^2) + i [(s \sm M_\b^2)\,M_\a\, \G_\a + (s \sm M_\a^2)\,
M_\b\, \G_\b ]}\;,
\label{11}
\eeq

\ni
where $\G_{\a,\b}$ and $\g_{\a,\b}$ are obtained by using the parameters suited for each 
resonance in eqs.(\ref{9}) and (\ref{10}).\\[1mm]
$\bullet$ There is no simple relationship between the unitarized amplitude for the coupled system
and Breit-Wigner expressions for individual resonances.
The price to be paid for representing each resonance by a Breit-Wigner is a very high one,
since one will have to deal with coefficients and phases that depend strongly on the variable $s$.

\vspace{2mm}

An extended version of this talk, given at the XI International Conference on Hadron Spectroscopy, 
Rio de Janeiro, Brazil, August 2005, has been submitted for publication elsewhere.

\vspace{-2mm}


\end{document}